\documentclass{nature_withfigs}

\usepackage{amsmath,amsfonts,amssymb}
\usepackage{wrapfig}
\usepackage{graphicx}
\usepackage{bbm}
\usepackage{color}

\usepackage[english]{babel}
\makeatletter

\usepackage{babel}
\makeatother

\def\be{\begin{eqnarray}}
\def\ee{\end{eqnarray}}
\def\bee{\begin{eqnarray*}}
\def\eee{\end{eqnarray*}}

\title{Measurement-based quantum computation}

\author{H. J. Briegel$^{1, 2}$, D. E. Browne$^3$, W. D\"ur$^{1, 2}$, R. Raussendorf$^4$ and M. Van den Nest$^{2,5}$}

\begin{document}

\maketitle

\begin{affiliations}
\item 
Institute for Theoretical Physics, University of Innsbruck, Technikerstra\ss e 25, A-6020 Innsbruck, Austria
\item Institute for Quantum Optics and Quantum
Information of the Austrian Academy of Sciences, Technikerstra\ss e 21a, A-6020 Innsbruck, Austria
\item 
Department of Physics and Astronomy, University College London, Gower Street, London WC1E 6BT, UK
\item
University of British Columbia, Department of Physics and Astronomy, 6224 Agricultural Rd., Vancouver, BC V6T 1Z1, Canada
\item
Max-Planck-Institut f\"ur Quantenoptik, Hans-Kopfermann-Str. 1, D-85748 Garching, Germany
\end{affiliations}

\def\makeheadbox{}

\begin{abstract}
Quantum computation offers a promising new kind of information processing, where the non-classical features of quantum mechanics can be harnessed and exploited. A number of models of quantum computation exist, including the now well-studied quantum circuit model. Although these models have been shown to be formally equivalent, their underlying elementary concepts and the requirements for their practical realization can differ significantly. The new paradigm of measurement-based quantum computation, where the processing of quantum information takes place by rounds of simple measurements on qubits prepared in a highly entangled state, is particularly exciting in this regard. In this article we discuss a number of recent developments in measurement-based quantum computation in both fundamental and practical issues, in particular regarding the power of quantum computation, the protection against noise (fault tolerance) and steps toward experimental realization. Moreover, we highlight a number of surprising connections between  this field and other branches of physics and mathematics.
\end{abstract}

\maketitle

\section{Introduction}

Quantum computation is a promising and fruitful area of research, and impressive theoretical and experimental achievements have been reported in recent years.  At the same time, many fundamental questions remain unanswered.
Realizing a large-scale computational device with the technology available in the foreseeable future remains a challenge, and the full range of applications for a working quantum computer is still unknown. A number of quantum algorithms are known for particular problems, including factoring and simulation of other quantum systems, but discovering new quantum algorithms that outperform classical ones remains a great challenge. On a more fundamental level, we are still missing a good understanding of where the border in computational power between the classical and the quantum lies.

Even the very notion of what makes a quantum computer and what it should be capable of doing, is not entirely understood. The latter point is highlighted by the existence of different models for quantum computation, including the quantum circuit (or network) model \cite{De89, De89b}, adiabatic quantum computation \cite{Fa01}, the quantum Turing machine \cite{De85}, and measurement-based models such as teleportation-based approaches \cite{Go99, klm, Ni03, Ni03a} as well as the one-way quantum computer \cite{Ra01, Ra01a, Ra03a}. 
 Indeed, there seem to be many ways to exploit Nature for quantum information processing.

As the features of these models differ significantly, some computational schemes may lend themselves more than others to understand certain aspects of quantum computation and to overcome challenges in their experimental realization. The new paradigm of 
{\em ``measurement-based quantum computation''} (MQC) \cite{Go99, klm, Ni03, Ni03a,Ra01,Ra01a,Ra03a,Ali04,Pe04,Dan04,Chi05,PEPS2,Brow07,Gr06}, with the ``one-way quantum computer'' and the teleportation-based model as the most prominent examples,
is particularly promising in these respects, and provides a new conceptual framework in which these experimental and theoretical challenges can be faced. While in, e.g., the circuit model quantum information is processed by coherent unitary evolutions (quantum gates), in MQC the processing of quantum information takes place by performing sequences of adaptive 
measurements. Moreover, whereas in the teleportation-based model joint (i.e. entangling) measurements are used, in the one-way quantum computer---which will be the focus of this article---universal quantum computation can be achieved with \emph{single-qubit measurements only}.

\begin{figure}[ht]
\begin{center}
 \includegraphics[width=8cm, clip]{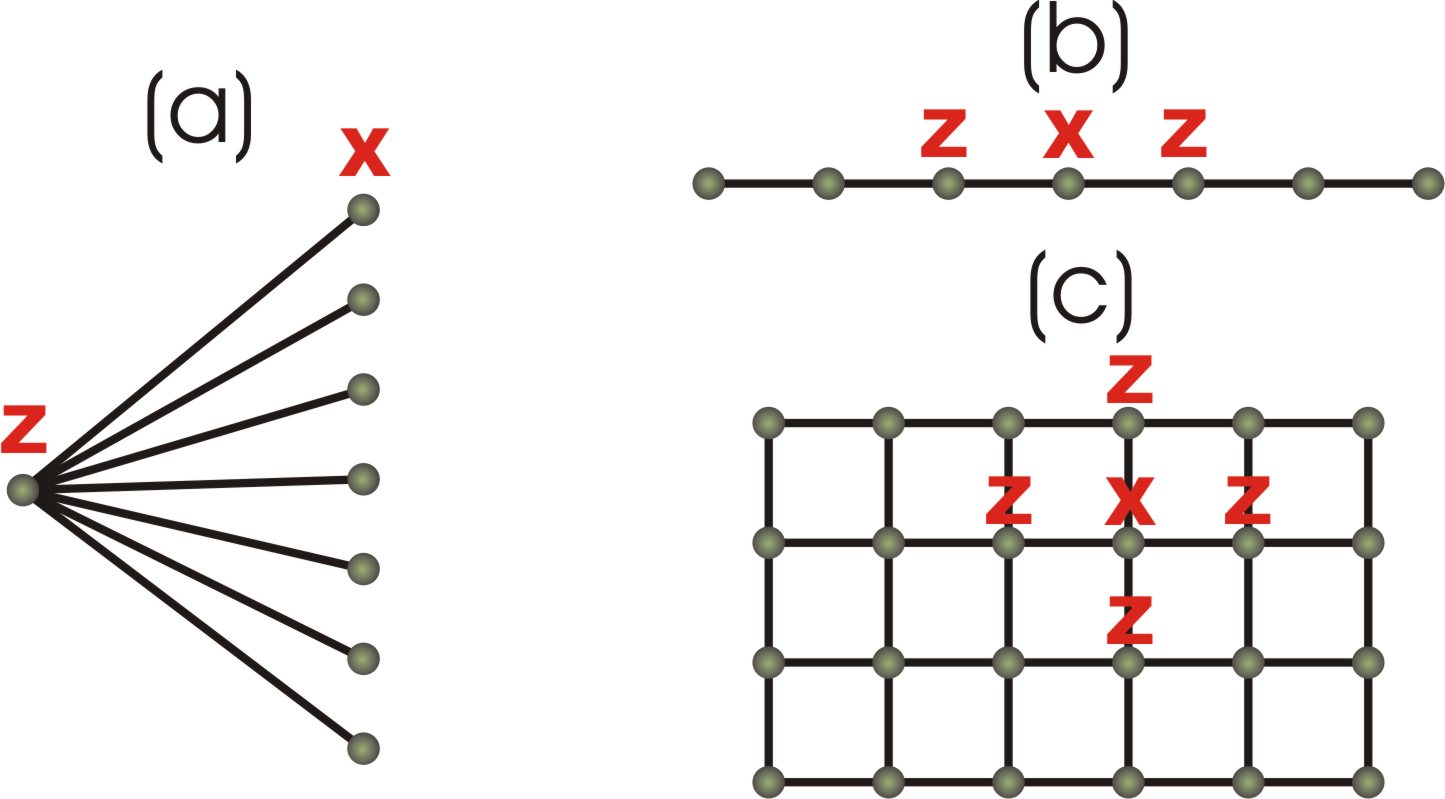}
\caption[]{\label{FigGS}
 {\bf BOX 1: Graph states.---} The cluster state \cite{Br01} belongs to a family of highly entangled multi-particle quantum states, which can be efficiently parameterized by mathematical graphs. These are the so-called ``graph states'' \cite{He06}. A  graph $G=(V,E)$ is a set of $N$ vertices ($V$), together with a set of edges $E\subseteq [V]^2 $, which connect the vertices in an arbitrary way. To every such graph we associate a specific $N$-qubit quantum state $|G\rangle$. The graph state $|G\rangle$ is obtained  by preparing all $N$ qubits in the $+1$ eigenstate of $\sigma_x$, namely $|+\rangle = (|0\rangle+|1\rangle)/\sqrt{2}$, and by applying two-qubit phase gates $U_{\mbox{\scriptsize{PG}}}=\mbox{ diag}(1,1,1,-1)$ between all pairs of qubits connected by an edge: $|G\rangle=\prod_{\{i,j\}\in E} U_{\mbox{\scriptsize{PG}}}^{(i,j)} |+\rangle^{\otimes N}$. Equivalently, $|G\rangle$ can be defined as a simultaneous fixed point of the correlation operators $K_j=\sigma_x^{(j)}\bigotimes_{\{i,j\} \in E} \sigma_z^{(i)}$, which are entirely determined by the graph.  The graphs in the figure correspond to (a) a Greenberger-Horne-Zeilinger (GHZ) state, 
 (b) a 1D cluster state and (c) a 2D cluster state, the latter being a universal resource for MQC. }
\end{center}
\end{figure}

\begin{figure}[ht]
\vspace*{-1cm}
 \begin{minipage}[c][1\totalheight]{4.25cm}%
 \includegraphics[width=7cm]{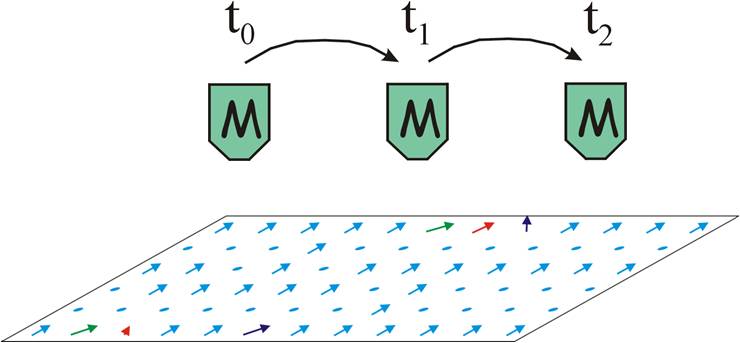} %
\put(-180,100){(a)}
\end{minipage}%
\hspace{5cm}\begin{minipage}[c][1\totalheight]{4.25cm}%
 \includegraphics[width=7cm]{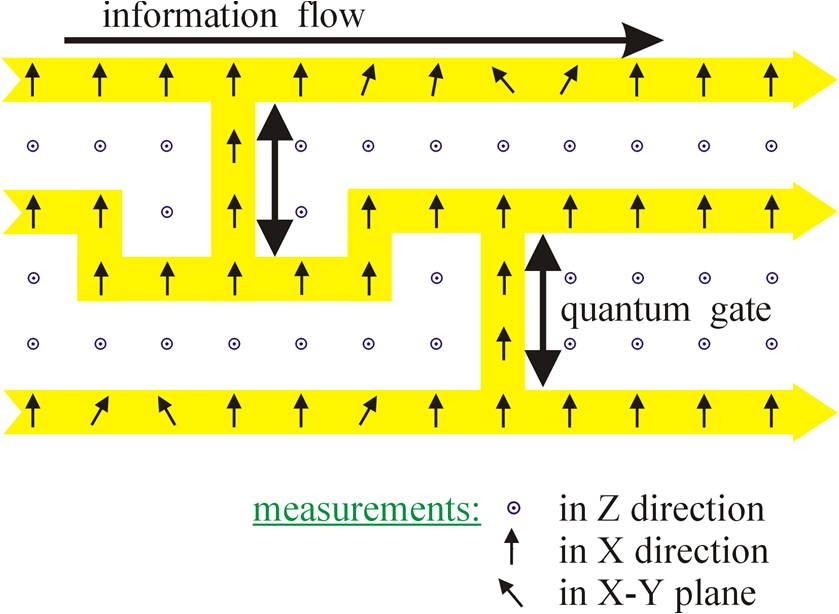} %
\put(-225,120){(b)}
\end{minipage}%

\caption[]{\label{FigMQC}{\bf BOX 2: The one-way quantum computer.---} In contrast to  the quantum circuit model, where quantum computations are implemented by unitary operations, in the one-way quantum computer, information is processed by sequences of \emph{single-qubit measurements}  \cite{Ra01}. These measurements are performed on a universal resource state---the 2D-cluster state \cite{Br01}---which does not depend on the algorithm to be implemented. A one-way quantum computation proceeds as follows (see (a) and (b)):

(i) A classical input is provided which specifies the data and the program.
\hspace{0.3cm} (ii) A 2D-cluster state $|{\cal C}\rangle$ of sufficiently large size is prepared. The cluster state serves as the resource for the computation.
\hspace{0.3cm} (iii) A sequence of adaptive one-qubit measurements $M$ (see (a)) is implemented on certain qubits in the cluster. In each step of the computation, the measurement bases (see (b)) depend on the program and on the outcomes of previous measurements. A simple classical computer is used to compute which measurement directions have to be chosen in every step.
\hspace{0.3cm} (iv) After the measurements, the state of the system has the form $|\xi^{\alpha}\rangle |\psi^{\alpha}_{\mbox{\scriptsize out}}\rangle$, where $\alpha$ indexes the collection of measurement outcomes of the different branches of the computation. The states $|\psi^{\alpha}_{\mbox{\scriptsize out}}\rangle$ in all branches are equal to the desired output state  up to  a local (Pauli)  operation; the measured qubits are in a product state $|\xi^{\alpha}\rangle$ which also depends on the measurement outcomes.
\hspace{0.3cm} \emph{The one-way quantum computer is universal}: even though the results of the measurements in every step of the computation are random, any quantum computation can deterministically be realized.

Notice that the temporal ordering of the measurements plays an important role and has been formalized e.g. in \cite{Ra01a,Dan04,Brow07}. For different perspectives and recent reviews on MQC, we refer the reader e.g. to \cite{Ni05, Jo05, BrBr06, Pe04, PEPS2}. \hfill $\square$

}
\end{figure}

More specifically, in one-way quantum computation the system is first prepared in a highly entangled quantum state, the \emph{2D-cluster state} \cite{Br01} (see Fig. \ref{FigGS}), independently of the quantum algorithm which is to be implemented---one thus calls the cluster state a ``universal resource''. In a second step, the qubits in the system are measured {\em individually}, in a certain order and basis---and it is this measurement pattern which specifies the entire algorithm (see Fig. \ref{FigMQC}). The quantum algorithm thereby corresponds, in an explicit sense, to a processing of quantum correlations.

Note that the one-way quantum computer is equipped with a remarkable feature, namely that the entire resource for the computation is provided by the entangled cluster state in which the system is initialized. This implies, in particular, that the computational power of such a quantum computer can be traced back entirely to the properties of its entangled resource state, thereby offering a focused way of thinking about the nature and strength of quantum computation. Moreover, the problem of an experimental realization of a quantum computer is now reduced to the preparation of a specific multi-particle state, and the ability to perform single-qubit measurements, offering practical advantages for certain physical set-ups. Finally, a fruitful marriage of ideas from MQC and topological error correction was recently achieved, paving the ground towards a scalable computational device that operates in a noisy environment.

The computational scheme of the one-way quantum computer was introduced in \cite{Ra01}. This work has stimulated numerous researchers, both theorists and experimentalists, to investigate MQC. Apart from offering an alternative approach towards realizing quantum computation, today MQC has become an interdisciplinary field of research, relating to entanglement theory, graph theory, topology, computational complexity, logic and statistical physics. It is the aim of this article to discuss a selection of recent results in MQC which illustrate the vigour and diversity of research in this field.

In this article, MQC will be considered in the sense of the one-way quantum computer; but we emphasize that there are other measurement-based approaches to quantum computation, as cited above.

\section{Experimental proposals and achievements}\label{Experiments}

\begin{figure}[t]
\begin{minipage}[c][1\totalheight]{4.25cm}%
 \includegraphics[width=7cm]{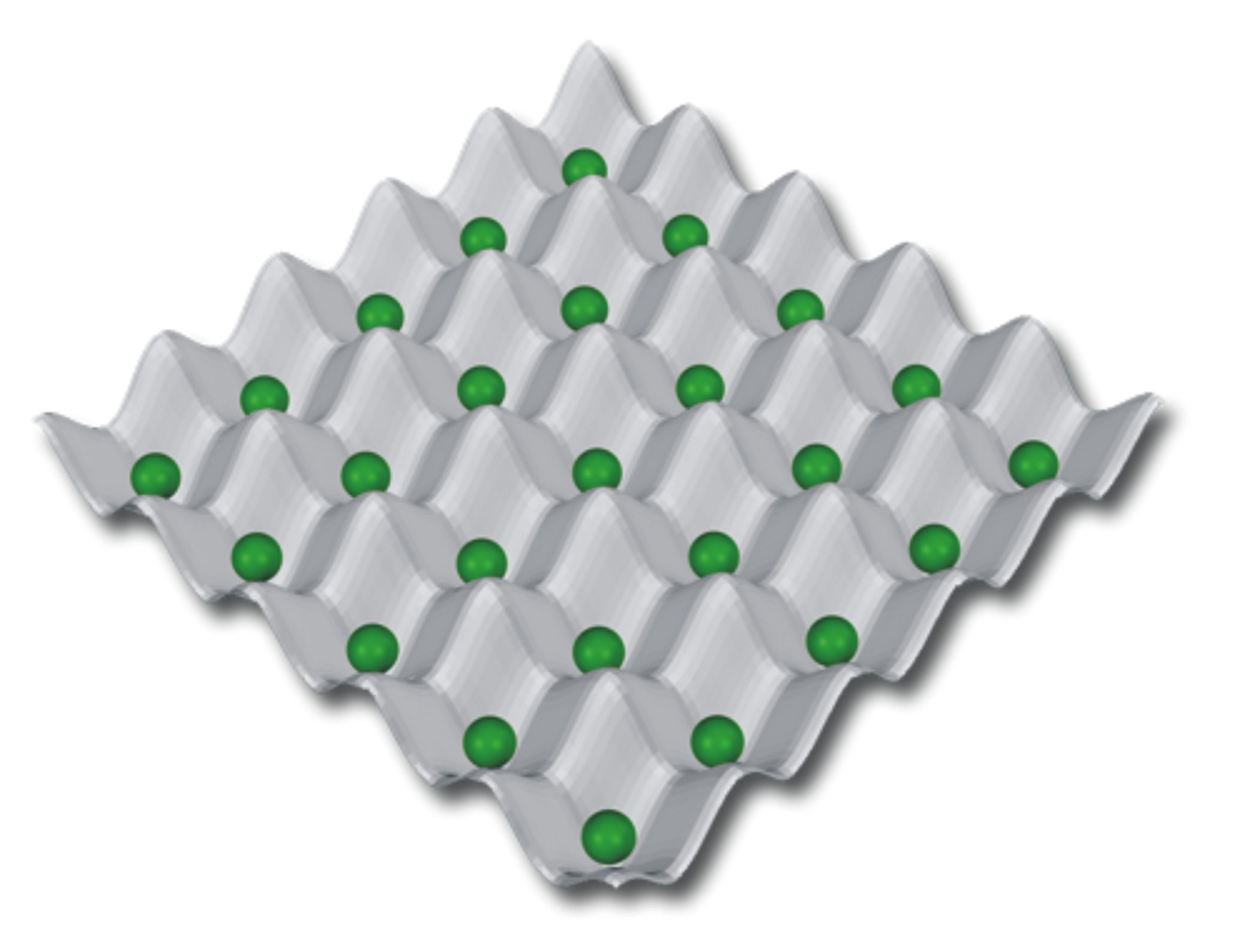} %
\put(-180,130){(a)}
\end{minipage}%
\hspace{5cm}\begin{minipage}[c][1\totalheight]{4.25cm}%
 \includegraphics[width=7cm]{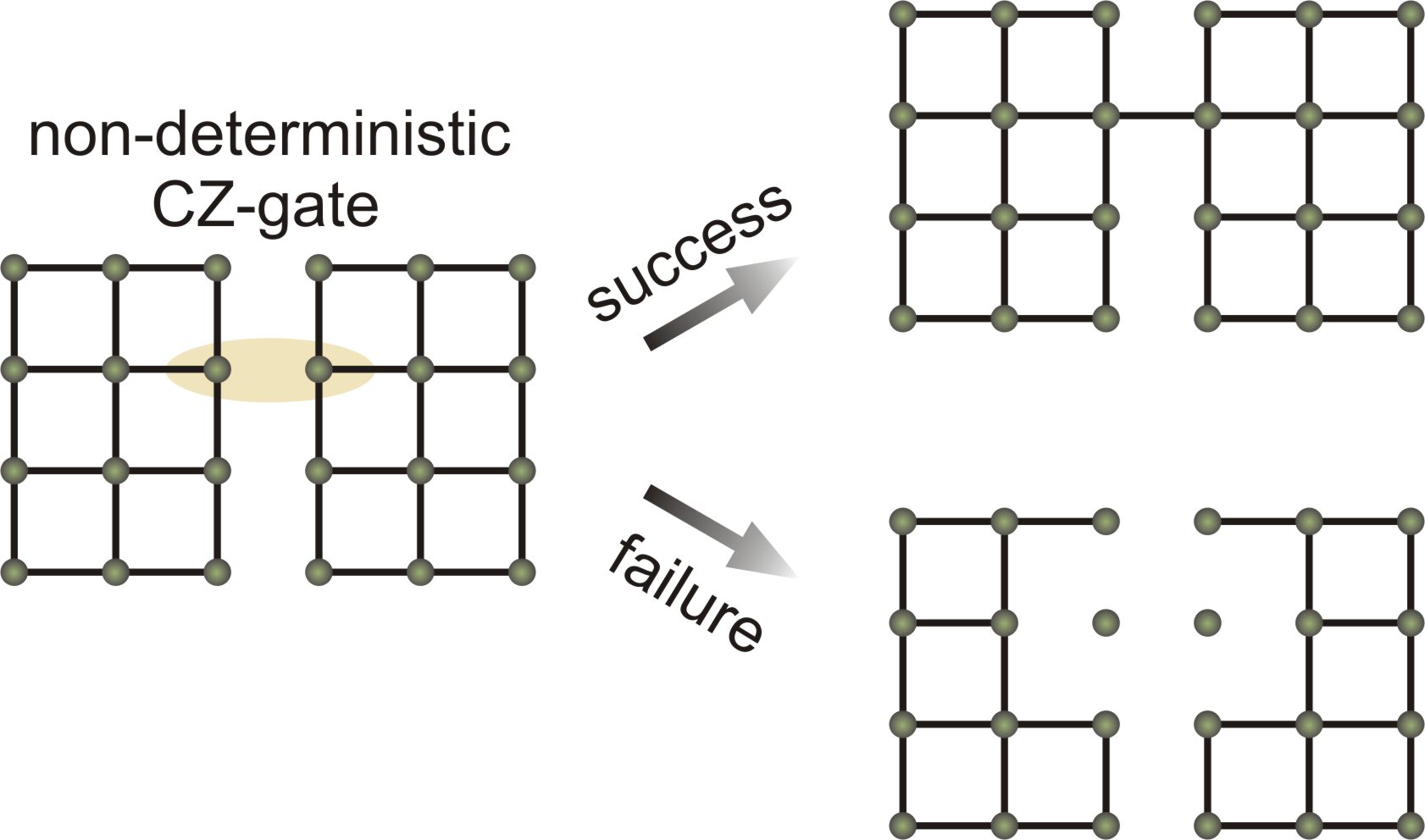} %
\put(-180,120){(b)}
\end{minipage}%

\caption[]{\label{Fig_Optical lattices} One-way computation can be implemented in a variety of systems, offering e.g. massively parallel operations, such as with atoms in optical lattices, or even non-deterministic operations, such as with photons. a) Cold neutral atoms are trapped in an optical lattice,
a standing wave potential formed from counter-propagating laser fields which holds atoms
in a three-dimensional array (reproduced, with permission, from I. Bloch, Nature Physics vol. 1, page 28, 2005).
State-dependent entangling operations can be realised in parallel across the lattice, by tuning
the trapping fields \cite{jakschbriegel99}. As was experimentally demonstrated in Ref. \cite{mandelcold},
this can be employed to generate, in few steps, a cluster state over the entire lattice (see also \cite{BrRaSc02}).
If the problem of addressing single lattice sites could be solved (see discussion in the main text),
this would open the way for a large-scale implementation of one-way quantum computation.

b) One-way quantum computation enables scalable quantum computation also with non-deterministic entangling logic gates. With probability $p<1$ the gate is achieved as required, but otherwise the participating qubits are measured. In a circuit based approach, this would destroy the coherence of the computation state, disrupting the computation. Cluster states can, however, be efficiently built with such operations, since a successful entangling operation increases the number of cluster state qubits.  A failure, while reducing the number of entangled qubits,  leaves the remaining qubits  in an intact cluster state. Strategies can be adopted to allow the efficient generation of cluster states of any size.}
\end{figure}

Apart from its useful conceptual status as an alternative model of quantum computation (see Sec. \ref{universality} and \ref{StatMech}), MQC can have practical advantages over the standard circuit model in a variety of different physical settings, from optical lattices and single photons to spatially separated matter qubits. %
%
In an optical lattice, cold atoms are kept in a standing-wave potential created by counter-propagating laser fields. The potential minima create a lattice of sites in which individual atoms can be trapped, storing quantum information in their long-lived internal states (see Fig \ref{Fig_Optical lattices}a). Tuning the polarization of the trapping lasers can induce entangling interactions \cite{jakschbriegel99, Ja05}
between neighboring atoms across the array and create a cluster state across the whole lattice \cite{Br01,BrRaSc02,mandelcold}.
In recent years there has been huge experimental progress in the trapping, cooling and manipulation of ultra-cold atomic gases
in optical lattices in one, two and three dimensions. For a recent review, see e.g. \cite{blochreview}. In particular, the creation of
a Mott insulator state with a crystal-like arrangement of single atoms in the lattice~\cite{Gr02,Ja98}, and the realization of controlled
entangling collisions \cite{mandelcold,jakschbriegel99} have been milestones for the coherent control of matter on the atomic level in these systems.
Recent experiments use exchange interactions in double-well potentials to create arrays of robust Bell pairs, which
could be used as an alternative way to create cluster states in the lattice\cite{Anderlini07, Bloch07} (see also \cite{Vau07}).

A remaining obstacle to the implementation of MQC in these systems is that the lattice spacing is typically of the order of the wavelength of the trapping light, too small for individual atoms to be addressed and measured. However, recent progress in the creation of lattices with wider spacing \cite{weisslattice}, sorting atoms in periodic potentials \cite{Mir06}, proposed methods for single-site addressing in tighter lattices \cite{optlatticespacing1}, as well as new methods achieving sub-wavelength resolution \cite{Go07,Daley07} are promising. Combining single-site addressing and lattice-wide entangling operations would allow large-scale one-way quantum computation to be realized in an optical lattice system.

Proof-of-principle experiments of a one-way quantum computer with few qubits have already been performed in which qubits are represented by photons. A photon can encode a qubit in, for example,  its polarization, or spatial degree of freedom. The generation and detection of single photons is making great advances and single-qubit operations can be achieved with precision via interferometers or polarization wave-plates. However, the deterministic two-qubit gates which would enable universal quantum computation cannot be achieved with interferometric techniques (linear optics) alone. By adding photon counters to interferometric networks non-deterministic entangling gates can, however, be achieved \cite{klm}. For some measurement-outcomes the gate is successful, for other "failure outcomes" the photon states are measured. In a standard quantum circuit these failure outcomes would be very damaging, since the measurement would destroy the coherence of the state, disrupting the computation.

The one-way quantum computation model provides an efficient way to enable scalable quantum computation with such non-deterministic gates \cite{nielsen, Ba05, Du05, Gro06,Ki07}, too.
The non-deterministic gates are used to create the cluster state, which can be done {\em off-line} and stochastically (see Fig. \ref{Fig_Optical lattices}b). Once the cluster state has been created, one can then proceed {\em deterministically} with the one-way quantum computation. Furthermore,  a polarizing beam splitter provides a simple non-deterministic entangling gate for efficient cluster state generation \cite{brownerudolph}.

For the above linear optical schemes to be truly scalable, extremely high single photon detection and generation efficiencies are required, beyond the capabilities of current experiments. However, demonstration experiments of the key components have been achieved, using post-selection \cite{zeilingercluster1, weinfurter, pan, zeilingercluster2, Val07, Ch07, Val07b}. Data is only kept when every detector fires, allowing loss or inefficiency errors to be discounted. These achievements have demonstrated the basic principles of one-way quantum computation, with recent experiments including simple algorithms and active feed forward of measurement results \cite{zeilingercluster2, Val07b}.

In addition to all-optical approaches, hybrid optical-matter schemes \cite{Ba05} to computation seem increasingly promising. In these schemes, matter qubits, such
as atoms, quantum dots or diamond NV centers, are kept isolated from one another, e.g. in separate cavities. Entangling operations can be achieved non-deterministically, via the emission of photons, entangled with the qubits, and then adopting non-deterministic linear optical entangling gates
\cite{moehringnature,La07}. Due to the non-deterministic nature of the entangling gates, the one-way model is, again, the most natural approach to scalable quantum computation either via non-deterministic strategies \cite{Kie07b,Du05}, repeat-until-success strategies \cite{Lim05}  or so-called ``broker-client'' approaches\cite{broker}.
Such approaches carry the significant advantages; individual qubits are isolated from each other, reducing correlated error, and the modularity of the approach facilitates scaling up to many qubits.

Scaling up these approaches, would appear to require optical networks of increasingly complicated switching circuits -- a potential barrier to their implementation.
However, by exploiting a phenomenon known as percolation, scalable quantum computation with non-deterministic gates is possible even with simple non-switching optical circuits \cite{Ki07}.
Once again, the simplicity and uniformity of the entanglement structure of cluster / graph states \cite{He06} (see Fig. \ref{FigGS}) is essential for this approach to succeed.

In addition to these works, there have been a series of proposals how to implement one-way quantum computation (or to create cluster states) in solid state systems, using superconducting circuits~\cite{Tan06} and quantum dots \cite{Bor05,Wei05}.

\section{Topological protection of information and fault-tolerant computation}\label{faulttolerance}

\begin{figure}[t]
  \begin{center}
   \includegraphics[width=8.6cm]{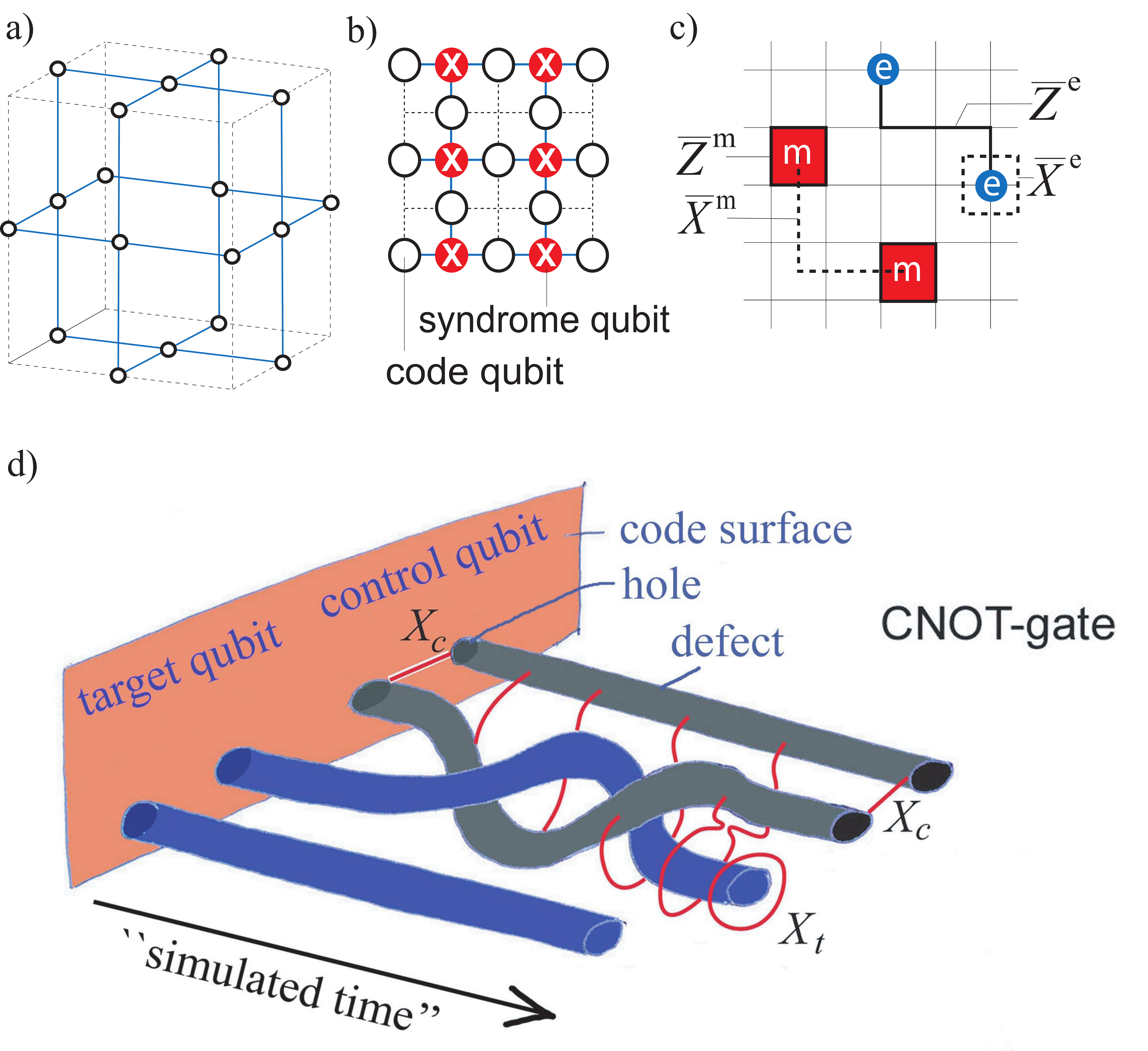}
   \caption{\label{CNOTexpl}Topological fault-tolerance from 3D cluster states.
     a) Elementary cell of the cluster lattice. b) A single 2D layer of the
     cluster. If the syndrome qubits are measured in the $X$-basis the code
     qubits are projected into a surface code \cite{Kit2} (code lattice indicated
     by dashed lines). c) A surface code with electric and magnetic holes,
     pairwise forming encoded electric and magnetic qubits, respectively.
     Strings supporting the  encoded Pauli operators are also shown.
     d) 3D cluster for a CNOT-gate between two encoded qubits. The gate is
     implemented by a monodromy between worldlines of holes in the code
     surface. The holes evolve in ``simulated
    time'' (the third cluster dimension). Also shown is the string
    corresponding to an encoded Pauli operator $\overline{X}$ on the control
    qubit and its evolution from the initial to the final codes surface.}
  \end{center}
\end{figure}

In realistic physical systems, decoherence
tends to make quantum systems behave more classically. One could therefore expect that decoherence would threaten any computational advantage possessed by a quantum computer. However, the effects
of decoherence can be
counteracted by quantum error correction \cite{ShorE}. In fact, arbitrarily
large quantum computations can be performed with arbitrary accuracy,
provided the error level of the elementary components of the quantum
computer is below a certain threshold. This important result is called the threshold theorem
of quantum computation
\cite{TT1,TT2,TT3,TT4}

Recent work has been dedicated to bringing fault-tolerance closer
to experimental reality. Proofs of fault-tolerance have been derived for
error-models showing the characteristic features of realistic physical
systems such as long-range correlated errors \cite{CoE}. Moreover, a
very high threshold of 3\% (on
average, one gate in thirty is allowed to fail) has been obtained for
a method using off-line preparation and post-selection within the
circuit model \cite{Kn2}.

What is the status of fault-tolerance in MQC ? Results have so far been obtained for the one-way
quantum computer. The existence of a non-zero threshold was
first proven by reduction to the circuit model \cite{RRth, ND, LA}.
Subsequent developments evolved along two lines. First, after it was
realized \cite{nielsen, brownerudolph, MY} that the one-way quantum computer may be advantageously
combined with the KLM scheme of optical quantum
computation \cite{klm} and fault-tolerant schemes using photons were
developed \cite {OCn,OCn2,LT2}. The dominant sources of
error in this setting are photon loss and gate inaccuracies.  The
constraint of short-range interaction and arrangement of qubits in a
2D lattice---a characteristic feature of the initial one-way quantum computer---is not
relevant for photons. In \cite{OCn}, both photon loss and gate
inaccuracies were taken into account yielding a trade-off curve
between the two respective thresholds. Fault-tolerant optical
computation is possible for e.g. a gate error rate of $10^{-4}$ and
photon loss rate of $3\times 10^{-3}$. In \cite{LT2} the stability
against the main error source of photon loss was discussed. With
non-unit efficiencies $\eta_S$ and $\eta_D$ of photon creation and
detection being the only imperfections, the very high threshold of
$\eta_S\,\eta_D>2/3$ was established \cite{LT}.

A second line of research \cite{RHG05} kept the geometric constraint of
nearest-neighbor interaction, which is a realistic scenario for stationary
qubits. To
achieve fault-tolerance it is then advantageous to increase the
lattice dimension from two to three. A 3D cluster
state combines the universality already found in the 2D counterpart
with the topological error-correction capabilities of the toric code
\cite{Kit1}. Error-correction is directly built into the cluster
lattice and yields a threshold of $6.7\times 10^{-3}$ \cite{RHG07},
for a model with probabilistic gate errors, including imperfect preparation of the cluster state.

Topologically protected quantum gates are performed  by measuring some regions of qubits in the $Z$-basis, which effectively removes the qubits from the state. The remaining cluster, whose qubits are measured in the $X$- and $X\pm Y$-basis, thereby attains a non-trivial topology in which fault-tolerant quantum gates can be encoded. A topological method of fault-tolerance can then be achieved \cite{RH06}. We shall explain this
method in some detail. First we map the 3D cluster state to a surface code \cite{Kit2} propagating in time. We consider a 3D cluster with elementary cell as in Fig.~\ref{CNOTexpl}a, and single out one spatial direction on the cluster as `simulated time'. As a first step, we consider a perpendicular 2D slice of this cluster, as shown in Fig.~\ref{CNOTexpl}b. The qubits are subdivided into code and syndrome qubits. As can be easily verified, measurement of the syndrome qubits in the $X$-basis projects the code qubits into a surface code state. In a 3D cluster consisting of many linked 2D slices, measurement of the code qubits results in teleportation of the encoded state from one slice to the next (plus local Hadamard gates), and measurement of the syndrome qubits amounts to measurement of the surface code stabilizer.

Now note that we can modify the code surface of Fig.~\ref{CNOTexpl}a by changing measured observables from $X$ to $Z$. For example, if we measure a syndrome qubit on plaquette $p$ in the $Z$ basis, we chose to not read out the surface code stabilizer $B_p=\bigotimes_{f\in \partial p}Z_f$ of the corresponding plaquette $p$. If we measure a code qubit on edge $e$ in the $Z$-basis, then we destroy syndrome information. Specifically, the eigenvalues of the surface code stabilizers $A_s=\bigotimes_{f|s \in \partial f}X_f$ on sites adjacent to $e$ will become undefined. These elementary operations lead to techniques for manipulating the code surface. First, we can punch holes into it. Holes come in two types, electric and magnetic. An electric hole is a site $s$ where the condition $\langle A_s\rangle=1$ is {\em{not imposed}} on the code space. Two electric holes support an electric encoded qubit. Similarly, a magnetic hole is a plaquette $p$ with $\langle B_p\rangle=1$ {\em{not imposed}} on the code space. Two magnetic holes support a magnetic qubit. Now, the encoded Pauli operators $\overline{Z}_e$, $\overline{X}_e$ and $\overline{Z}_m$, $\overline{X}_m$ for these encoded qubits can be described in geometric terms. Namely, (E) for the electric qubit on $\{s,s^\prime\}$, $\overline{Z}_e$ is a tensor product of $Z$'s along a string stretching from $s$ to $s^\prime$. $\overline{X}_e$ is a tensor product of $X$'s along a string looping around either $s$ or $s^\prime$. (M) for the magnetic qubit on $\{p,p^\prime\}$ its just the same with the roles of $Z$ and $X$ interchanged; See Fig.~\ref{CNOTexpl}c. To protect these encoded qubits against harmful errors, the holes are enlarged from one site or plaquette to extended connected sets of sites or plaquettes.

By slightly shifting the locations of $Z$-measurements from one cluster slice to the next the holes in the code surface can be moved and fused. This gives rise to encoded unitary gates and measurements, respectively. As an example, consider the topological encoded CNOT-gate displayed in Fig.\ref{CNOTexpl}d, between a magnetic control and electric target qubit. It is realized by `moving' one of the electric holes around one of the magnetic holes. Again, by `moving' we mean slightly shifting the locations of $Z$-measurements in consecutive cluster slices. It is important to note that when the holes are moved, the above rules (E),(M) continue to apply. Thus, the strings corresponding to the encoded Pauli operators are dragged along. We can now easily verify the four conjugation relations for the CNOT gate in a topological manner, by sliding the operator strings forward along the hole world-lines. One example, $X_c \longrightarrow X_c\otimes X_t$ ($t$: target, electric; c: control, magnetic) is displayed in Fig.~\ref{CNOTexpl}d, the other three are similar. Note that fault-tolerant CNOT gates between two electric or two magnetic encoded qubits can also be performed but are more complicated; they require fusion of holes \cite{RHG07}.

Finally, by way of the described mapping between the 3D cluster state and a 2D surface code changing in time, there exist a circuit variant of fault-tolerant cluster-state computation which only requires a two-dimensional lattice of qubits. Also this 2D variant only requires translation-invariant
nearest-neighbor interaction. It yields a threshold value
of $7.5 \times 10^{-3}$ \cite{RH06} which is the highest known threshold
for a two-dimensional local architecture (see also \cite{Sv06}).
This scenario is suited for realization in e.g. optical lattices, but also arrays of superconducting qubits, and ion traps.


\section{Entanglement as a resource for computational power}\label{universality}

{\it Universality.---} In MQC, universal quantum computation is realized by performing sequences of single-qubit measurements on a system which has initially been prepared in a 2D-cluster state. As individual measurements can only destroy entanglement, the entire computational power of the one-way quantum computer is carried by the entanglement structure of its resource state. Can we understand and quantify what are the essential features which make the 2D-cluster state, and possibly other states, ``universal resources'', i.e. capable of enabling arbitrary measurement-based quantum computations? Here we report recent progress on this issue and discuss some important outstanding questions.

A first feature which is to be emphasized, is that there exist several natural notions of ``universality'' \cite{Va06b}.
In its strongest form, universality is defined as the capability of generating every possible quantum state from the resource by means of single-qubit operations.
As an important example, the 2D-cluster state is a universal resource in this sense. Thus, a universal measurement-based quantum computer is then identified as a device which allows for universal state preparation by local operations only \cite{Va06a,Va06b}. This implies that, whenever any given type of entanglement---as quantified by an appropriate entanglement measure \cite{Vi00}---is to be generated from the resource state, it must already be present in the resource itself, as single-qubit operations cannot add entanglement to the system. Using this intuition, it can be shown that every such universal resource must be \emph{maximally entangled} with respect to all types of entanglement \cite{Va06a,Va06b}. The 2D-cluster state provides a key
example of such a maximally entangled resource state---but we emphasize that the entanglement criteria hold for every possible universal state preparator.

This insight can be utilized to develop a systematic framework to investigate which states are universal state preparator resources for MQC, and in particular to obtain no-go results. Following this approach, it can e.g. be shown that $n$-particle 1D-cluster states, Greenberger-Horne-Zeilinger (GHZ) states, 
 W-states, 
 Dicke states, 
 and certain ground states of strongly correlated 1D spin systems, are not universal resources \cite{Va06b}. Note that many of the above states are considered to be highly entangled. However, in each case there is at least one type of entanglement which is non-maximal, implying that these states cannot be universal state preparator resources.

States which do not violate any of the entanglement criteria, include the graph states \cite{He06} associated with various types of regular 2D lattices (triangular, hexagonal, Kagome), as well as lattices with a high degree (up to about $~40\%$ corresponding to the classical site-percolation threshold in 2D)
of defects---and, in fact, it can be proven that such states are universal in the same way as the 2D-cluster states \cite{Va06b, Br07}.

The notion of ``universality'' in MQC used in the preceding discussion is the strongest one possible, and can be relaxed in several ways.
Most importantly, one may study an altogether different concept of universality, where the goal is \emph{not} to prepare arbitrary quantum states as outputs.
In contrast, it is only required that a universal measurement-based quantum computer is capable of (efficiently) reproducing the \emph{classical} output of any quantum computation implemented on a standard gate array quantum computer. While it seems difficult to formulate entanglement-based criteria for this form of universality, one can study it from a \emph{constructive} perspective: in \cite{Gr06,Gr07} it is shown how such universal resources for MQC---beyond the 2D-cluster states---can systematically be constructed. The underlying structure of this approach 
can be described mathematically using the language of matrix-product-states or projected-entangled-pairs \cite{PEPS2,MPS,PEPS}. The main conclusion of this investigation is that a number of extremal entanglement features (such as e.g. maximal localizable entanglement) exhibited by 2D-cluster states  no longer have to be present in universal resources if only classical outputs are considered \cite{Gr06,Gr07}. At present it is not clear whether the latter form of universality is fundamentally distinct from universal state preparation under relaxed conditions, such as encodings \cite{Va06b,Ba06,Ta07,Vau07}. This issue is presently under investigation.

{\it Classical simulation.---} When a state is identified as not being a universal resource, this does not necessarily mean that it could not be used for a specific quantum computational task for which it may still outperform classical computers. This naturally leads to a study of classical simulation of MQC, where we  ask: ``Which resource states do {\it not} offer any computational speed-up with respect to classical computation?'' This question is closely related to major investigations in condensed matter theory, where one studies under which conditions quantum systems can efficiently be described and simulated.

Recently, several techniques have been developed to tackle classical simulatability of MQC, and considerable progress has been reported \cite{Ma05, Jo06, Sh06, Yo06}. For example, for the majority of states which have been identified above as not being universal, 
one has shown  that efficient classical simulation is possible. More precisely, for such states it is possible to efficiently and exactly compute the outcome probabilities of any sequence of single-qubit measurements.  Further interesting examples of simulatable states include the toric code states \cite{Br06} and the ground states of several 1D spin systems \cite{MPSfaceful}. The techniques
invoked to obtain these results are again centered around entanglement; {for example, the entanglement measure ``entanglement width'' can be used} to identify efficiently simulatable
states \cite{Va06c,Sh06}.

In this section we have described two---in some sense complementary---investigations regarding the origin of the quantum computational power.  Even though these investigations were carried out within the specific framework of the one-way computer, the insights are of general relevance, since the different quantum computational models can simulate one another efficiently and are thus, in this complexity theoretic sense, equivalent.

Although significant progress has been obtained in the issues of universality and simulatability, these matters are far from being fully understood. Most importantly, it is at present not known whether a universal quantum computer disallows efficient classical simulation---in other words, whether quantum computers are truly (exponentially) ``more powerful'' than classical ones. Nevertheless, we believe that the recent works provide the first steps to understanding this important but difficult question.

\section{MQC and classical statistical mechanics}\label{StatMech}

We have already seen that the study of the principles of MQC turns out to be connected to different fields, e.g. entanglement theory, topology, and graph theory. In this section we show how some of the central questions raised in the study of MQC are related to notions of the statistical physics of \emph{classical} spin systems, such as the Ising model and the Potts model \cite{Wu84}. Such spin models were introduced in the context of (anti-)ferromagnetism, but they seem to have wide-spread applications not only in physics, but also e.g. in optimization theory and biology \cite{Mezard87}.

Let us illustrate these connections by considering the example of the Ising model in the presence of an external field. In the Ising model, one envisages a large lattice ${\cal L}$ (e.g., a 2D square lattice) of (classical) spins $s_a=\pm 1$. The spins interact according to the Hamiltonian function \be\label{Ising} H_{\cal L}(\{s_a\}) = - \sum_{\langle a, b\rangle} J_{ab}s_a s_b - \sum_{a} h_a s_a.\ee The couplings $J_{ab}$ and $h_a$ determine the strength of the pairwise interaction and the external field, respectively. The lattice ${\cal L}$ may be arbitrary, in the sense that lattices of arbitrary dimension---and in fact arbitrary graphs---are possible. The partition function $Z_{\cal L} = \sum_{\{s_a\}} e^{-H_{\cal L}(\{s_a\})/(k_b T)}$, where $k_b$ is the Boltzmann constant and $T$ the temperature, is a central quantity in this context, and from it other relevant system properties such as free energy or magnetization can be derived.

The connection between MQC and the Ising model is given by a mapping of the Ising model on an arbitrary lattice ${\cal L}$ to an MQC model where the entangled resource is determined by the geometry of ${\cal L}$, \be\label{partition_overlap} Z_{\cal L} \cong \langle \psi_{\cal L}|\bigotimes_{i}|\alpha_i\rangle. \ee This expression states that the partition function $Z_{\cal L}$ is identified with a quantum mechanical amplitude
which is obtained as an overlap between two quantum states \cite{Va07a, Br06, Va07} (see also \cite{Bo07}). The multiparticle entangled state $|\psi_{\cal L}\rangle$ \cite{Va07} encodes the interaction pattern and is a {\em graph state} \cite{He06} (see Fig. \ref{FigGS}). The product state $\bigotimes_{i}|\alpha_i\rangle$ contains no entanglement and specifies interaction strengths and local magnetic fields as well as the temperature of the model.

How does the expression (\ref{partition_overlap}) allow us to connect this model with MQC? To see this, simply consider an MQC with the state $|\psi_{\cal L}\rangle$ as a resource state. Then, according to (\ref{partition_overlap}),  the (computation of the) partition function corresponds to a specific \emph{measurement pattern} on the resource state $|\psi_{\cal L}\rangle$. In this way, a connection is drawn between concepts from MQC and statistical physics.

This simple connection opens the possibility to obtain a cross-fertilization between statistical mechanics and MQC.
For example, one finds a notable relation between the \emph{solvability} of the Ising model on a lattice ${\cal L}$ and the \emph{computational power} of an MQC operating on a resource state $|\psi_{\cal L}\rangle$. This brings us back to the issue of the power of quantum computation, and the (im)possibility of an efficient classical simulation. The central quantities to be considered in this context are overlaps between the resource state and product states: these quantities need to be computed to determine with which probabilities the outcomes of local measurements occur, and 
exactly those overlaps are identified with the partition function $Z_{\cal L}$. Eq. (\ref{partition_overlap}) now implies that any model where the partition function can efficiently be computed, leads to a corresponding MQC which offers no computational advantage over classical devices, and vice versa. For example, the solvability of the 2D Ising model without magnetic fields implies that MQC on the toric code state \cite{Kit1} can be efficiently simulated \cite{Br06}. In turn, the efficient classical simulation of MQC on stabilizer states with bounded tree-width \cite{Va06c}, as demonstrated in \cite{Sh06}, yields a novel classical algorithm to efficiently calculate the partition function on (inhomogenous) $q$-state spin models on tree-like-graphs
with or without magnetic fields
\cite{Va07a,Va07}.

The relation (\ref{partition_overlap}) can also be exploited in a different sense. 
For example, in \cite{Va07} the universality of the 2D-cluster states was used to prove that the 2D Ising model with magnetic fields is ``complete'' in the sense that the partition function of any $q$-state spin model on an arbitrary lattice can be expressed as a special instance of the partition function of the 2D Ising model (in a complex parameter regime).

We believe that several other interesting applications can be found. For one, these connections enable one to phrase statistical physics problems naturally in a quantum mechanical setting \cite{Br06,Va07,Va07a,Bo07,Lidar04,PEPS,Somma06,Aharanov07}. This may open a new path towards e.g. \emph{quantum algorithms} for problems in this area.

Finally, we mention that additional connections between MQC and other fields have been established, e.g. to decidability of formal languages in mathematical logic \cite{logic}.

\section{Outlook}\label{Outlook}

The discovery of the one-way quantum computer has opened up new experimental avenues toward the realization of quantum computation in the laboratory. At the same time it has challenged the traditional view of the very nature of quantum computation itself. 

For the future, we see several open problems and challenges. On the experimental side, one of the main challenges is to realize large scale quantum computation in the laboratory, beyond proof-of-principle demonstrations. For measurement-based schemes, we believe that optical lattices are still one of the most suitable candidates that allow us to create large-scale cluster states with high efficiency. However, the problem of addressing single sites in the lattice - and thus to fully implement one-way computation - remains unsolved to date, even though there are new and encouraging developments. Recent progress with photonic one-way quantum computation is exciting, and few-qubit applications e.g. in the context of the quantum repeater are conceivable; however, for a scalable setup that could go much beyond proof-of-principle experiments, new single photon sources with higher efficiency are needed. In the meantime, a variety of new and promising proposals for one-way quantum computation using hybrid systems have been put forward, which combine the advantages of different physical implementations. It remains to be seen which system will become practical in the long run. Apart from engineering issues, the capabilities of the system to naturally accommodate quantum error correction (i.e. on the hardware level), and thus to facilitate fault-tolerant operation, will play an important role.

On the theoretical side, the study of fault-tolerant schemes and the search for new quantum algorithms will remain central issues, related to the fundamental issue of universality, classical simulation, and the role of entanglement. In the context of measurement-based computation, a deeper understanding of universality (i.e. the properties that make an entangled state a universal resource) will help us to find resource states that are tailored to specific physical systems. It will also provide a basis to improve schemes for fault-tolerant computation, e.g. by choosing more robust states or by reducing the physical overhead. A deeper understanding of efficient classical simulation, on the other hand, will narrow down the set of interesting quantum algorithms. A promising strategy to find new quantum algorithms is to connect quantum computation with other fields. An example of such a connection -- with classical statistical mechanics -- has been given here, but there seem to be many more.

In conclusion, it seems that the conceptual framework of the one-way model continues to be an attractive alternative platform for experimental and theoretical investigations of quantum computation and its ramifications.

\begin{addendum}
 \item We like to thank Immanuel Bloch for helpful comments on the manuscript.
The work was supported by the European Union (QICS,OLAQUI,SCALA), EPSRC's QIPIRC programme and the Austrian Science
Foundation.
 \item[Competing Interests] The authors declare that they have no
competing financial interests.
 \item[Correspondence] Correspondence
should be addressed to H.J.B.~(email: hans.briegel@uibk.ac.at).
\end{addendum}

\end{document}